# External Control over Magnon-Magnon Coupling in a Two-Dimensional Array of Square Shaped Nanomagnets

Swapnil Barman[1,a], Pratap Kumar Pal[2,b,#] and Rajib Kumar Mitra[1,3,*]

[1]Technical Research Centre, S. N. Bose National Centre for Basic Sciences, JD Block, Sector III, Salt Lake City, Kolkata – 700106, India

[2]Department of Condensed Matter and Material Physics, S. N. Bose National Centre for Basic Sciences, JD Block, Sector III, Salt Lake City, Kolkata – 700106, India

[3]Department of Chemical and Biological Sciences, S. N. Bose National Centre for Basic Sciences, JD Block, Sector III, Salt Lake City, Kolkata – 700106, India

[a,b]These authors contributed equally to this work.

[*]Email: rajib@bose.res.in

[#]Email: palpratap07@gmail.com

## Abstract

The field of hybrid magnonics has gained significant momentum in recent years, driven by its potential to enable coherent information transfer and quantum transduction. This study delves into the tunable magnon-magnon coupling within a two-dimensional array of $Ni_{80}Fe_{20}$ (Permalloy) square nanomagnets by modulating its internal magnetic field configuration. Using a broadband ferromagnetic resonance (FMR) spectroscopy, we systematically investigate the influence of the orientation of external magnetic field and microwave power on the coupling properties. Our findings reveal a pronounced anticrossing behavior, indicative of robust magnon-magnon coupling, whose strength can be tuned through both static and dynamic external parameters. The ability to modulate coupling strengths in this system highlights its potential for developing flexible and adaptive magnonic devices, crucial for future applications in quantum information processing and spintronic technologies. This work not only broadens the understanding of magnon-magnon interactions in complex geometries but also opens new avenues for the design of next-generation quantum magnonic systems.

## 1. Introduction

During the past two decades, significant advancements in nanoscale magnonics have revealed their great potential for creating energy-efficient and high-speed on-chip data communication and processing technologies[1]. Nanoscale magnetic structures exhibit salient properties, including shorter spin waves (SWs)[2], device scalability[3], non-reciprocity[4], waveguiding[5], and hybridization[6], making them ideal for applications such as SW filters[7], transistors[8,9], logic gates[10], memory devices[11], and wave-based computing[12]. Despite these advantages, a major challenge remains in the relatively low conversion efficiencies associated with the excitation[13], manipulation, and detection of magnons[14]. Efficiently converting electrical or optical signals into SWs, controlling their propagation with minimal losses, and detecting output signals accurately are crucial for practical applications. These inefficiencies are mainly due to the energy dissipation during SW propagation and the limited sensitivity of detection methods, posing significant barriers to the widespread use of magnonic devices in high-performance information processing systems. Hybrid or quantum magnonics[15,16] have the potential to overcome this challenge. They focus on the coherent interactions of magnons with other platforms like photons[17–20], phonons[21–25], plasmons[26,27], excitons[28,29], and superconducting qubits[30–33]. These interactions enable the manipulation of a system's response in one degree of freedom through controlled excitations in another, leading to high-speed and energy-efficient operations. Hybrid quantum systems[15,16,34] provide a new framework for the coherent transfer of quantum states, advancing research in fundamental physics and practical applications in quantum computing, communication, and sensing[35–38]. The exploration of magnons in hybrid systems began with studies on spin ensembles coupled to microwave photons[39,40]. Magnetic materials, characterized by high spin densities, extended coherence time, and low damping[41,42], not only facilitate ultrastrong coupling but also serve as effective transducers between various platforms, a capability that is crucial for quantum information processing[19]. Recent perspectives[43,44] have highlighted the significant progress in controlling the magnon based hybrid quantum interactions in various kinds of systems starting from bulk crystals to single qubit. This coupling strength is notably higher than that of light-matter interactions due to the shared host medium, which resolves spatial mode overlap issues and ensures robust and efficient coupling[45].

Building on foundational research, extensive experimental efforts have been made to unravel the complexities of microwave cavity-enhanced magnon coupling across diverse systems.

Yttrium Iron Garnet (YIG) has emerged as the predominant material in cavity magnonics[1,46] due to its ultralow damping and high spin density[47]. However, recent studies have expanded the scope of magnon-magnon coupling, incorporating novel systems such as compensated ferrimagnets[48], two-dimensional antiferromagnetic materials like $CrCl_3$[49], synthetic antiferromagnetic structures[45,50,51], YIG/Co [52], and YIG/Py [53] heterostructures. While bulk and thin-film systems have significantly advanced the field, they face challenges in device miniaturization and on-chip integration, particularly due to their reliance on large magnetic volumes for strong coupling[54]. This limitation has driven a shift towards nanoscale hybrid magnonics, with a focus on enhancing miniaturization and compatibility with complementary metal–oxide–semiconductor (CMOS) technology. This transition is crucial for overcoming scalability issues and developing more compact, efficient magnonic devices. Recent literature has also made significant strides in understanding magnon-magnon coupling within a variety of non-ellipsoidal shaped magnetic nanostructures, including hexagonal dots[55,56], nanocross[57–59], diamond dots[60], and triangular dots[25]. These studies explored how geometric confinement of the nanoelements and lattice symmetry of the array influence the magnon-magnon coupling. Furthermore, various extrinsic and intrinsic means to tune the coupling parameters, including external magnetic field, microwave excitation power, geometry of the structure and material choices have been explored, providing insights into optimizing hybrid magnonic systems for specific applications. Despite the above efforts, the field of nanoscale magnonics is still at its infancy, which demands exploration of new structural geometry and external control for wider tunability of the coupled systems, leading towards applications in on-chip quantum communication and transduction.

In this study, we focus on the tunable magnon-magnon coupling in a two-dimensional array of square shaped $Ni_{80}Fe_{20}$ (Permalloy) nanomagnets arranged on a square lattice. The square shaped nanodots can show a range of continuously tunable internal field configuration as well as inter-dot interaction fields as a function of array geometry and bias magnetic field strength and orientation leading to reconfigurable magnonic properties[61,62]. However, it has not yet been explored in the context of magnon-magnon coupling. Here, we aim to fill the gap in the existing literature by the investigation of magnon-magnon coupling in this specific nanodot array using broadband ferromagnetic resonance (FMR) spectroscopy. Bias magnetic field dependent SW frequencies in this system revealed a clear anti-crossing between two SW modes indicating the existence of magnon-magnon coupling between those modes. Further, the coupling strength could be enhanced significantly by the bias field angle as well as the microwave excitation power. The dissipation rates of the modes and the cooperativity have been analyzed to obtain strong coupling between the modes. Micromagnetic simulations reproduced the observed behaviour qualitatively and the simulated SW mode profiles and magnetic field distribution throw key insights into the observed behaviour. We have performed further simulations to show the coherent propagation of the hybrid SW modes through the lattice when locally excited at the centre of the lattice.

## 2. Experimental and Computational Methods

### A. Sample Fabrication

A 20-nm-thick Py nanodot array consisting of square-shaped dots, organized in a rectangular lattice symmetry, was fabricated on top of a Si (100) substrate with a co-ordination of electron-beam lithography (EBL) and electron-beam evaporation (EBE) techniques[58]. The Py layer is deposited on the Si (100) substrate in a high vacuum chamber at a base pressure of $2 \times 10^{-8}$ Torr. The array spans a size of (20 μm × 200 μm), with the nanodots having width of 375 nm (±10 nm). The dots are spaced at an edge-to-edge interval of $S_x$ = 100 nm (±5 nm) horizontally and $S_y$ = 150 nm (±5 nm) vertically. For the broadband FMR measurements, a coplanar waveguide (CPW), made of Au, with 150 nm thickness, 300 μm length, 25 μm central conducting width (*w*), and 50 Ω characteristics impedance ($Z_0$) was fabricated on top of the square-shaped dot array. The CPW was patterned via maskless optical lithography, and is separated from the nanodot arrays by a 60-nm-thick insulating layer of $Al_2O_3$ deposited on top of the Py layer, made by EBE. Succeeding this, a 5-nm-thick Ti protective layer was deposited on top of the Au layer, both of which were done at a base pressure of $6 \times 10^{-7}$ Torr. The scanning electron microscope (SEM) image in Figure 1(b) shows the square-shaped nanodot array and the slight asymmetry along the y-axis in the elements along with rounded bottom-left corners. These imperfections were taken into account for the micromagnetic simulations.

### B. Measurement Techniques

The FMR spectroscopy to study the SW resonance and its magnetic-field dispersion was done using a high-frequency broadband vector network analyzer (VNA, Agilent PNA-L, model no.: N5230C, frequency range: 10 MHz to 50 GHz) [63], a home-built high-frequency probe station equipped with a non-magnetic ground-signal-ground (G-S-G) picoprobe (GGB Industries, model no.: 40A-GSG-150-EDP) and a coaxial cable. Microwave signal having power ranging from -15 dBm to +6 dBm was applied at a specified frequency range, to the CPW and the reflected signal was collected back to the VNA by the G-S-G probe connected with the co-axial cable. The SW spectra and its magnetic-field dispersion of the sample were obtained from the frequency-swept real part of the scattering (S)-parameter (Re($S_{11}$)) in the reflection geometry. The bias magnetic field ($H$) was systematically varied across the full range from -1.4 kOe to +1.4 kOe. However, the anticrossing phenomenon was specifically observed within a narrower field range of 0.2 kOe to 0.8 kOe, prompting us to focus our analysis on this region. We systematically adjusted the angular orientation ($\varphi$), but beyond 25°, the spectral features changed drastically, and the anticrossing behavior disappeared. Therefore, we limited our angular analysis to the range $0° \leq \varphi \leq 30°$. The spectra were measured at 20 Oe intervals and at room temperature. A schematic of the experimental geometry is shown in Figure 1(a). The measured SW field dispersion for the nanodot array with $P$ = -15 dBm at $\varphi$ = 0° is shown in Figure 1(d) with a single raw spectrum shown in Figure 1(c).

### C. Micromagnetic Simulations

Computational study via micromagnetic simulations was conducted on the nanodot array, using finite-difference method (FDM) based Mumax$^3$ [64] and OOMMF [65] software. Furthermore, a homebuilt code named DOTMAG[66] was used to calculate the SW mode profiles (power and phase) and the magnetic field distributions. The SEM image of the nanodot array was replicated, and a two-dimensional (2D) periodic boundary condition (PBC) was applied to take into account of the large-area array of nanodots used in the experiment. The array was discretized into identical rectangular cells of dimensions $4 \times 4 \times 20$ nm$^3$. The lateral sides of the cells were kept below the exchange length of Py (5.2 nm) for including the effect of Heisenberg exchange interaction in the SWs. The material parameters used in the simulation were gyromagnetic ratio ($\gamma$) = 17.95 MHz/Oe[25], saturation magnetization ($M_s$) = 780 emu/cm$^3$[67,68], Gilbert damping constant ($\alpha$) = 0.008[69] (for dynamic studies), exchange stiffness constant ($A_{ex}$) = $1.3 \times 10^{-6}$ erg/cm[70], and magnetic anisotropy field ($H_k$) = 0, for Py. During the 4 ns dynamic simulations, $m_z$ (out-of-plane magnetization) has been saved at intervals of 10 ps. The SW spectra at a particular bias magnetic fields are provided by the fast Fourier transform (FFT) of the time-resolved data.

## Results and Discussion

### Impact of Bias-field angle on Magnon-Magnon Coupling

The bias field dependence of the SW absorption spectra for the sample, within the range of -1.4 kOe $\leq H \leq$ +1.4 kOe, is plotted at intervals of 20 Oe, forming a surface plot at a fixed azimuthal angle $\varphi$ = 0° and a microwave power of $P$ = -15 dBm, as presented in Figure 1(d).

The real part of the forward scattering parameter $S_{11}$, representing the raw FMR spectrum for $H$ = 380 Oe, is shown in Figure 1(c). The magnetic field dispersion of SW modes in Figure 1(d) shows non-monotonic and asymmetric behaviour in positive and negative field regimes. While the positive field regime reveals the presence of three prominent SW modes, labelled as M*, M1, and M2, the negative field regime shows a dominant SW mode for majority of the field values along with a very weak mode. Between -0.4 kOe ≤ $H$ ≤ 0, blurred modes are observed in the surface plot due to large broadening of the modes. The positive field regime shows a prominent anticrossing phenomenon between two modes M1 and M2 even at $\varphi$ = 0° and $P$ = -15 dBm, indicating the existence of magnon-magnon coupling between those two modes. In Figure 1(c) the anticrossing gap is labelled as '2$g$', which is the spacing between the peaks of M1 and M2. Evidently, there is a transfer of mode power from one SW branch to the other and coexistence of both the SW over a certain field range, which is the characteristic feature of the avoided crossing or anticrossing. In this study, we have investigated how different external factors, such as the bias field angle and the amplitude of microwave excitation power, influence and modulate the observed anticrossing phenomenon.

First, we examine the impact of the azimuthal angle, $\varphi$ of the bias field on the SW mode anticrossing. As mentioned before the anticrossing disappeared at $\varphi$ = 30°, and we have limited our investigation only up to 30°. Figure 2(a)-(c) displays the surface plots of the bias-field dispersion of SW frequencies at $\varphi$ = 0°, 15° and 30°, respectively at $P$ = -15 dBm. The surface plots of the intermediate angles can be found in Figure S2(a)-(f) of the Supplementary Information, which clearly showcases the continuous variation of the anticrossing gap. The anticrossing regions, indicated by white dashed lines in Figures 2(a) and (b), were used to determine the magnon-magnon coupling strength ($g$). These values were obtained by fitting the SW spectra with a dual-peak Lorentzian function, as shown in Figure 2(d), for various $\varphi$ values. The peak-to-peak values of the coupled modes from the fits correspond to "2$g$" values (in GHz) in all cases. The extracted $g$ values are subsequently plotted as a function of $\varphi$ in Figure 3(a), revealing an increase from 208 MHz at $\varphi$ = 0° to 305 MHz at $\varphi$ = 25°.

Table 1: Experimental and simulated values of $g$, $k_1$, $k_2$, and $C$ from fittings at different values of $\varphi$.

| **$\varphi$ in °** | **$g$ in GHz** | | **$k_1$ in GHz** | | **$k_2$ in GHz** | | **$C$** | | **Coupling Regime** |
|---|---|---|---|---|---|---|---|---|---|
| | **Expt.** | **Sim.** | **Expt.** | **Sim.** | **Expt.** | **Sim.** | **Expt.** | **Sim.** | |
| 0 | 0.2065 | 0.325 | 0.07165 | 0.2165 | 0.147 | 0.197 | 4.04 | 2.47 | Strong (as $g > k_1$, $k_2$) |
| 5 | 0.195 | 0.345 | 0.071 | 0.198 | 0.154 | 0.14 | 3.47 | 4.29 | |
| 10 | 0.205 | 0.345 | 0.0675 | 0.195 | 0.135 | 0.115 | 4.61 | 5.30 | |
| 15 | 0.211 | 0.35 | 0.1125 | 0.21 | 0.155 | 0.23 | 2.55 | 2.53 | |
| 20 | 0.23 | 0.34 | 0.17 | 0.175 | 0.117 | 0.135 | 2.65 | 4.89 | |
| 25 | 0.306 | 0.37 | 0.21 | 0.65 | 0.105 | 0.135 | 4.24 | 1.56 | Strong (Expt.)/Intermediate (Sim.) |

Additionally, we extracted the dissipation rates $k_1$ an $k_2$ from the half width at half maximum (HWHM) of the coupled SW modes from the fits and calculated the cooperativity factor ($C = g^2/k_1k_2$) and the values of $g$, $k_1$ and $k_2$ have been listed in Table 1 to evaluate the evolution of the nature of the coupling as a function of $\varphi$. It shows that the coupling remains in the strong coupling regime for all the angular orientations. To obtain further insights of this magnon-magnon coupling, we simulated the SW response of the studied sample using Mumax$^3$ as described above for $0° \leq \varphi \leq 25°$, as shown in Figure 2(d). The simulations also revealed the anticrossing between modes M1 and M2 validating the observed behaviour and its origin in the internal magnetic field configuration. Moreover, the simulation captures the increasing trend in the coupling strength with $\varphi$ similar to the obtained experimental results. The simulated $g$ values, plotted as a function of $\varphi$ in Figure 3(b), qualitatively match the experimental trend well, though the precise quantitative agreement is not obtained. This discrepancy may arise from challenges in accurately incorporating edge roughness, deformation, and surface textures in FDM-based simulations, as detailed in the literature[71]. Furthermore, other factors, such as the simulations being performed at T = 0 K while the experiments are conducted at room temperature, also contribute to the differences observed. The simulated values of $g$, $k_1$, $k_2$ and $C$ have also been listed in Table 1 for a comparison of the coupling behaviour with the experiment as a function of $\varphi$. We observed that the coupling predominantly remains in the strong coupling regime, as demonstrated experimentally. However, at 25°, the system transitions to the intermediate coupling regime, characterized by $k_2<g<k_1$. This discrepancy is most likely due to the differences between the experimental and simulation conditions as described above but it is a rarity.

**Impact of Microwave Power on Magnon-Magnon Coupling**

Building on our analysis of the impact of bias field angle on SW mode anticrossing and the magnon-magnon coupling, we explored the possibility of further tuning the coupling using another external parameter, i.e. the microwave excitation power. Recognizing that the maximum coupling strength was achieved at $\varphi$ = 25°, we maintained this orientation and systematically increased the microwave power from -15 dBm to the maximum attainable power of +6 dBm in our experimental setup, to monitoring the corresponding modulation in the SW absorption spectra. Figure 4(a) illustrates the real part of the $S_{11}$ absorption spectra at different power levels, revealing a clear modulation in the behavior of the coupled SW modes. To quantify this effect, we calculated the coupling strength $g$ as a function of microwave power, as shown in Figure 4(c). The data exhibit a pronounced upward trend, with the coupling strength rising from approximately 305 MHz at -15 dBm to 319 MHz at +6 dBm.

To further substantiate our experimental findings, we performed micromagnetic simulations to model the SW response of the sample across varying microwave power levels. Figure 4(b) illustrates the simulated FFT power versus frequency plots at different microwave powers, providing a comparative perspective alongside our experimental data. The simulations reveal a trend consistent with our experimental observations: an increase in peak-to-peak frequency gap, denoted as "2g" in GHz, with higher microwave power. Figure 4(d) presents the extracted $g$ from simulations as a function of $P$, which qualitatively mirrors the experimental trend. The simulations confirm that the coupling strength between SW modes increases with rising microwave power, thus corroborating the experimental evidence. Despite the qualitative agreement, minor quantitative discrepancies were noted between the simulations and the experimental results due to the reasons discussed previously. The values of $g$, $k_1$, $k_2$, and $C$ as obtained from the experiment and simulation are listed in Table 2. The results show that the system consistently remains in the strong coupling regime across all measured power levels, as $g > k_1, k_2$ in all cases. The increase in microwave excitation power results in a noticeable enhancement of the coupling strength in both experimental and simulated data. This demonstrates that microwave power modulation is an effective tool for fine-tuning magnon-magnon coupling. The ability to maintain strong coupling across varying power levels suggests the potential for designing robust magnonic systems with customizable properties, enhancing coherence times and information preservation over extended spatial distances.

Table 2: Experimental and simulated values of $g$, $k_1$, $k_2$, and $C$ at different values of $P$.

| $P$ in dBm | $g$ in GHz | | $k_1$ in GHz | | $k_2$ in GHz | | $C$ | | Coupling Regime |
|---|---|---|---|---|---|---|---|---|---|
| | Expt. | Sim. | Expt. | Sim. | Expt. | Sim. | Expt. | Sim. | |
| -15 | 0.305 | 0.37 | 0.21 | 0.175 | 0.105 | 0.065 | 4.21 | 12.04 | Strong (as $g > k_1, k_2$) |
| -6 | 0.3075 | 0.47 | 0.155 | 0.175 | 0.088 | 0.155 | 6.93 | 8.14 | |
| 0 | 0.3135 | 0.52 | 0.189 | 0.180 | 0.109 | 0.240 | 4.77 | 6.26 | |
| +6 | 0.319 | 0.585 | 0.1775 | 0.180 | 0.177 | 0.170 | 3.23 | 11.18 | |

## SW Mode Profiles Analysis

To further elucidate the nature of the observed SW modes, we conducted an in-depth analysis of the spatial power and phase maps of these modes generated using DOTMAG. Figure 5 presents the power and phase maps of the two interacting SW modes giving rise to mode anticrossing, analyzed both at and away from the anticrossing field, for different values of $\varphi$. These SW modes exhibit mixed standing wave characteristics, denoted by quantization numbers [m, n], where m and n correspond to the quantization numbers in the backward volume (BV) and Damon-Eshbach (DE) geometry, respectively. Away from the anticrossing point, mode 1 (M1) and mode 2 (M2) display distinct quantization patterns, with M1 having lower quantization numbers [3,1] than M2 [7,1]. As these two SW modes interact, they undergo energy transfer and eventually converge to exhibit identical behavior with the same quantization number [5,1] for the hybrid mode at the anticrossing region at $\varphi = 0°$. As the bias field orientation is systematically increased, the axis of mode quantization rotates accordingly and SW modes become more complex. However, one or both quantization numbers of the hybrid SW modes remain identical indicating the strong magnon-magnon coupling and energy transfer between the modes.

To further understand the variation in coupling properties, we conducted micromagnetic simulations to analyze the magnetostatic field distribution of the sample at varying $\varphi$, focusing specifically on the two extreme cases, i.e. $\varphi = 0°$ and $\varphi = 25°$. Figures 6(a) and (b) present these distributions, where the contrast in color between the two angles highlights a significant reduction in uncompensated magnetic moments as $\varphi$ increases from 0° to 25°. This reduction is indicative of the enhanced static dipolar interactions between neighboring nanomagnets, consistent with findings from previous studies[58]. To further quantify this effect, we performed line scans across the stray-field distributions between consecutive nanodots, as depicted by the white dashed lines in Figures 6(a) and 6(b). The full distribution of these stray-field values is detailed in Figure S3 (a) of the Supplementary Information. From this analysis, we extracted the stray-field values ($H_s$) at each angle $\varphi$, which are plotted in Figure 6(c). The data reveals a substantial increase in $H_s$, from negligibly small value at $\varphi = 0°$ to about 259 Oe at $\varphi = 25°$. This pronounced enhancement in the stray field suggests that the interaction between unsaturated spins at the edges of adjacent nanodots plays a crucial role in modifying the SW dynamics. Specifically, the increased static dipolar interaction, mediated by these stray fields, appears to be a key factor contributing to the observed variation in magnon–magnon coupling within this system as was observed for nanodots of other shapes[57–59].

In order to understand the coherent propagation of hybrid modes, we have performed further numerical simulations. Figure 7 demonstrates an investigation of the above at a bias field of 0.46 kOe for two distinct in-plane orientations $\varphi = 0°$ and 25°, where two extreme values of coupling strength have been observed. This behavior is elucidated through the spatial power distribution of the hybrid SW modes. In the simulation, we have applied a time-dependent "sinc" field with a 20 GHz cut-off frequency to a small square region (100 nm × 100 nm) at the center of the nanodot array. At a microwave power of −15 dBm and $\varphi = 0°$, both the modes propagate efficiently along the vertical direction of the array but along the horizontal direction, they cease to propagate beyond the nearest neighbour, showing the limitation in the coherent propagation of hybrid mode for this configuration. However, when the in-plane angle is increased to $\varphi = 25°$ (with $P = -15$ dBm), the two hybrid modes propagate uniformly across the entire array, showing their superior efficacy for the coherent information propagation using hybrid magnonics in nanoscale systems. When the microwave power if increased to +6 dBm at the same bias field orientation, i.e. $\varphi = 25°$ the effect is further amplified promising coherent information transfer over longer distance (although not shown here due to limitation in the size of the array studied). These pronounced enhancements in coupling between the two magnon modes at higher bias field angle and microwave excitation power underscore the role of dynamic dipolar interactions within the nanodot array. These interactions intensify the magnon-magnon coupling, leading to the observed increase in the anticrossing gap. These findings underscore the potential to manipulate the propagation characteristics of the SW modes in such systems through external control of both the magnetic field orientation and excitation power amplitude. The outcome of this work promotes the square nanomagnet arrays as a potent platform for offering and modulating the hybrid magnons using external parameters and stimuli leading towards the development of on-chip quantum information processing and quantum transduction devices using magnons.

## 3. Conclusion

In summary, we investigated the spin wave (SW) dynamics in square-shaped nanomagnets made of permalloy dispersed in a rectangular lattice using broadband ferromagnetic resonance (FMR) spectroscopy. A key finding of our research is the observation of magnon-magnon coupling, manifested as an anticrossing feature in the bias-field dispersion of two interacting SW modes. Recognizing the importance of coupling properties in evaluating the performance of hybrid quantum systems, we explored the potential for tuning this coupling through external parameters. A systematic variation of the bias magnetic field orientation revealed a marked increase in the anticrossing gap with the bias-field angle. Specifically, the coupling strength, quantified by the parameter $g$, increased from 208 MHz at $\varphi = 0°$ to 305 MHz at $\varphi = 25°$. We calculated the dissipation factors and the cooperativity, confirming that the hybrid system falls within the 'strong coupling' regime. This variation in $g$ is attributed to strengthening of dipolar interactions between adjacent nanodots, a conclusion supported by micromagnetic simulations that, despite minor quantitative differences, are qualitatively aligned with our experimental observations. Further, we examined the influence of microwave power on the coupling strength at a fixed angular orientation ($\varphi = 25°$), where coupling strength was found to be maximum. Notably, as the power was raised from -15 dBm to +6 dBm, the coupling strength exhibited a systematic increase, reaching up to 319 MHz. This dependency on microwave power further highlights the role of dynamic dipolar interactions in modulating magnonic properties, suggesting promising avenues for the development of adaptable hybrid magnonic devices. Analysis of SW mode profiles at and away from the anticrossing field revealed that the hybrid modes attain identical mode profiles, a feature that substantiates the formation of magnon-magnon hybridization. The observed SW mode profiles and magnetostatic-field distributions underscore the importance of inter-element dipolar interactions, mediated by stray fields, in governing magnon-magnon coupling variations. Hence, our work offers critical insights into the mechanisms driving magnon-magnon coupling in patterned nanomagnetic arrays. The demonstrated ability to control this coupling through external magnetic fields and microwave power paves the way for the design of advanced hybrid magnonic devices, suitable for quantum transduction and quantum information processing. This tunability not only enhances the functional versatility of magnonic systems but also sets the stage for future exploration in the burgeoning field of hybrid and quantum magnonics.

## Acknowledgements

S.B. acknowledges Technical Research Centre (TRC), S. N. Bose National Centre for Basic Sciences for funding. P.K.P. acknowledges SNBNCBS for Bridge fellowship. We sincerely thank Dr. Yoshichika Otani for the sample.

## Conflict of Interest

The authors declare no conflict of interest.

**Figures:**

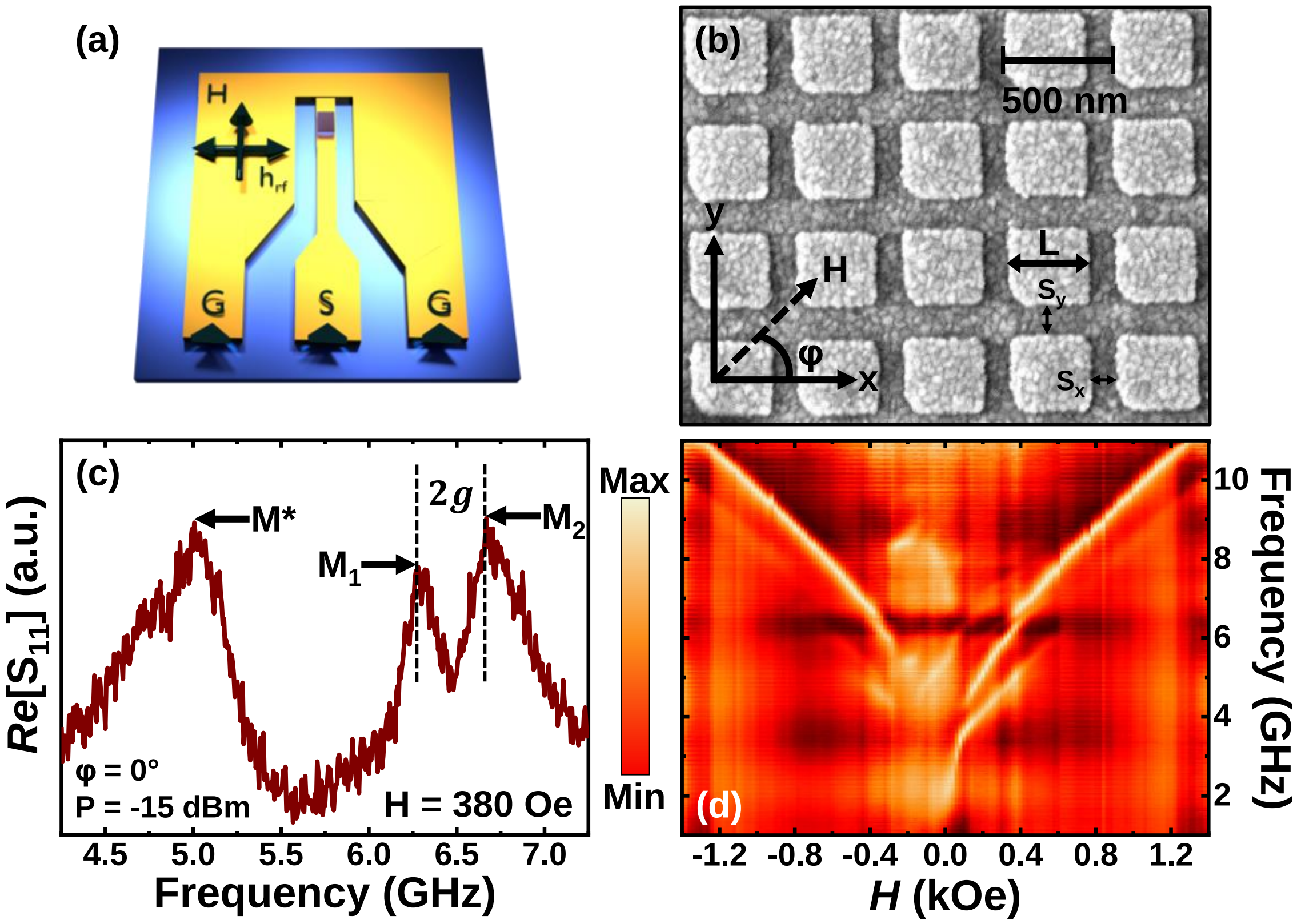

**Figure 1.** (a) Illustration of the experimental setup used for broadband ferromagnetic resonance (FMR) measurements. (b) Scanning electron microscope (SEM) image depicting the array of Permalloy ($Ni_{80}Fe_{20}$) square nanodots with a side length of approximately 330 ± 10 nm. The dots are spaced at intervals of 100 ± 5 nm horizontally ($S_x$) and 150 ± 5 nm vertically ($S_y$), with a thickness of about 20 nm. The orientation of the applied magnetic field is indicated for clarity. (c) Raw FMR spectrum, represented by the real part of the $S_{11}$ parameter, are shown at a magnetic field of 380 Oe for an in-plane angle $\varphi = 0°$, measured at a power level of −15 dBm. (d) A surface plot generated from the FMR spectra collected at varying magnetic field $H$ is presented, with the corresponding color map provided at the left side of the figure.

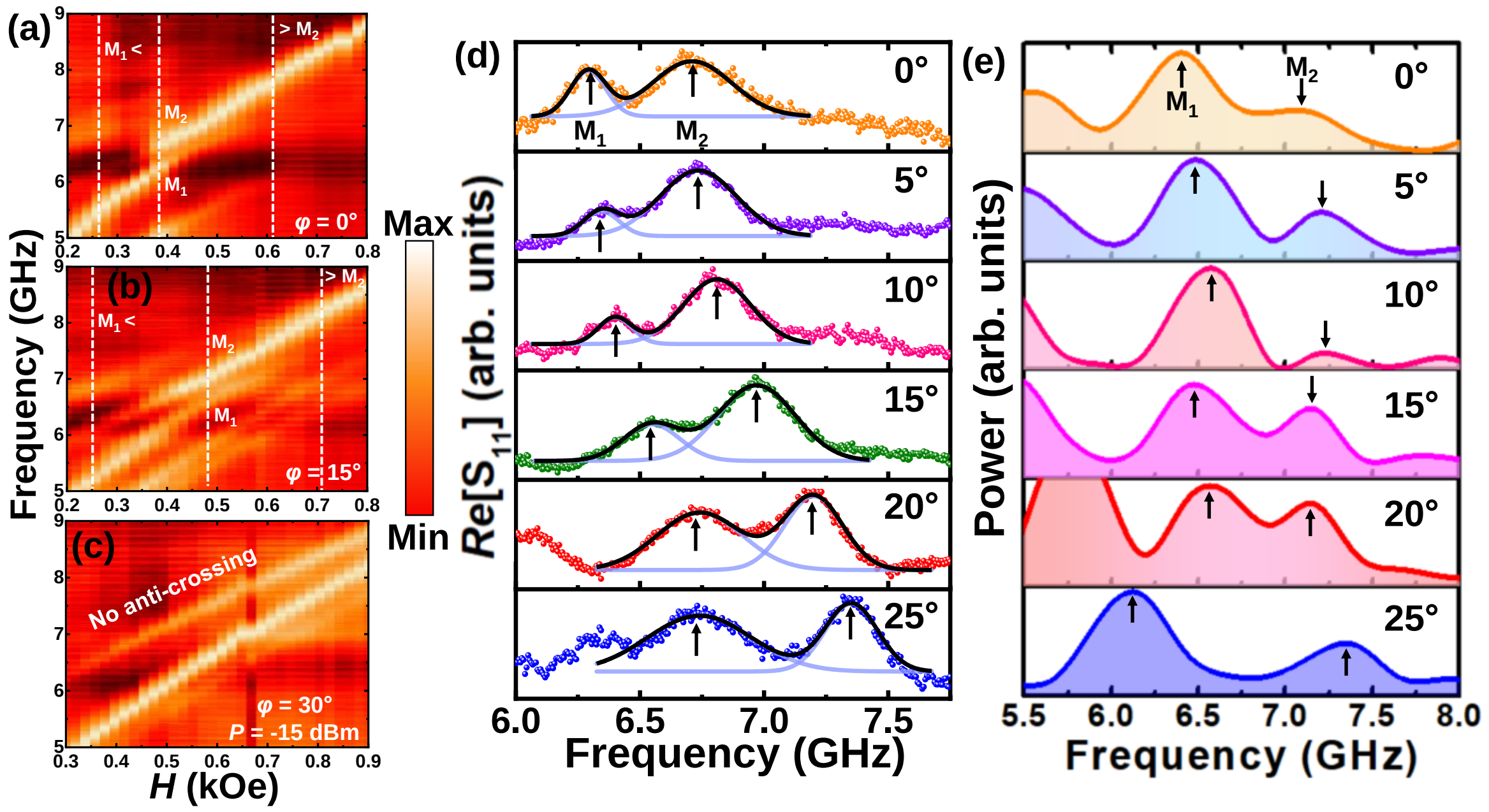

**Figure 2.** Surface plots illustrating the dependence of SW mode frequencies on the applied bias magnetic field for the square dot array at a power level of −15 dBm. The plots correspond to in-plane angles of (a) $\varphi$ = 0°, (b) $\varphi$ = 15° and (c) $\varphi$ = 30°. The anticrossing region, along with two additional points distant from this region, are indicated by white dashed lines. (d) The real component of the $S_{11}$ parameter is shown as a function of frequency within the anticrossing region for various $\varphi$ values. The solid black lines depict the theoretical fits, and the SW modes are marked as M1 and M2, indicated by upward arrows. (e) Simulated SW spectra corresponding to the given $\varphi$ values are presented for comparison.

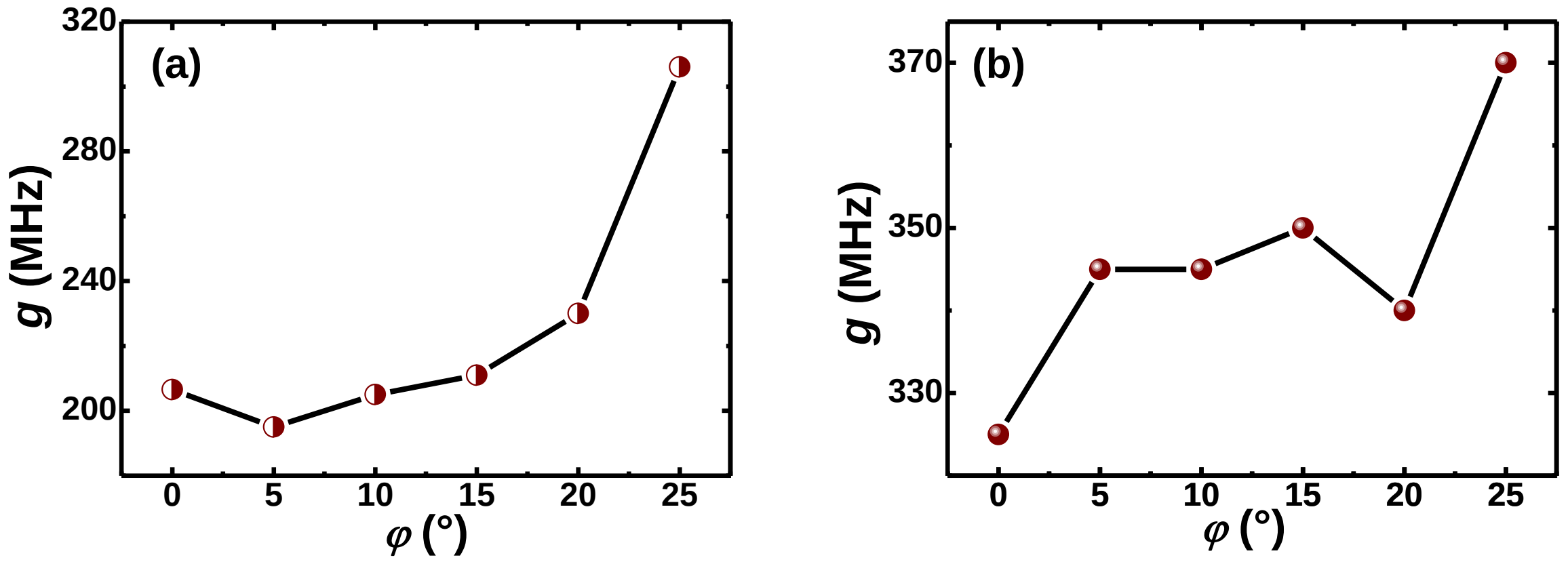

**Figure 3.** (a) Experimental measurements and (b) corresponding simulations depicting the variations of the coupling strength $g$ as a function of the in-plane angle $\varphi$. The data points are illustrated using symbols, while the continuous lines connect these symbols to guide the eye.

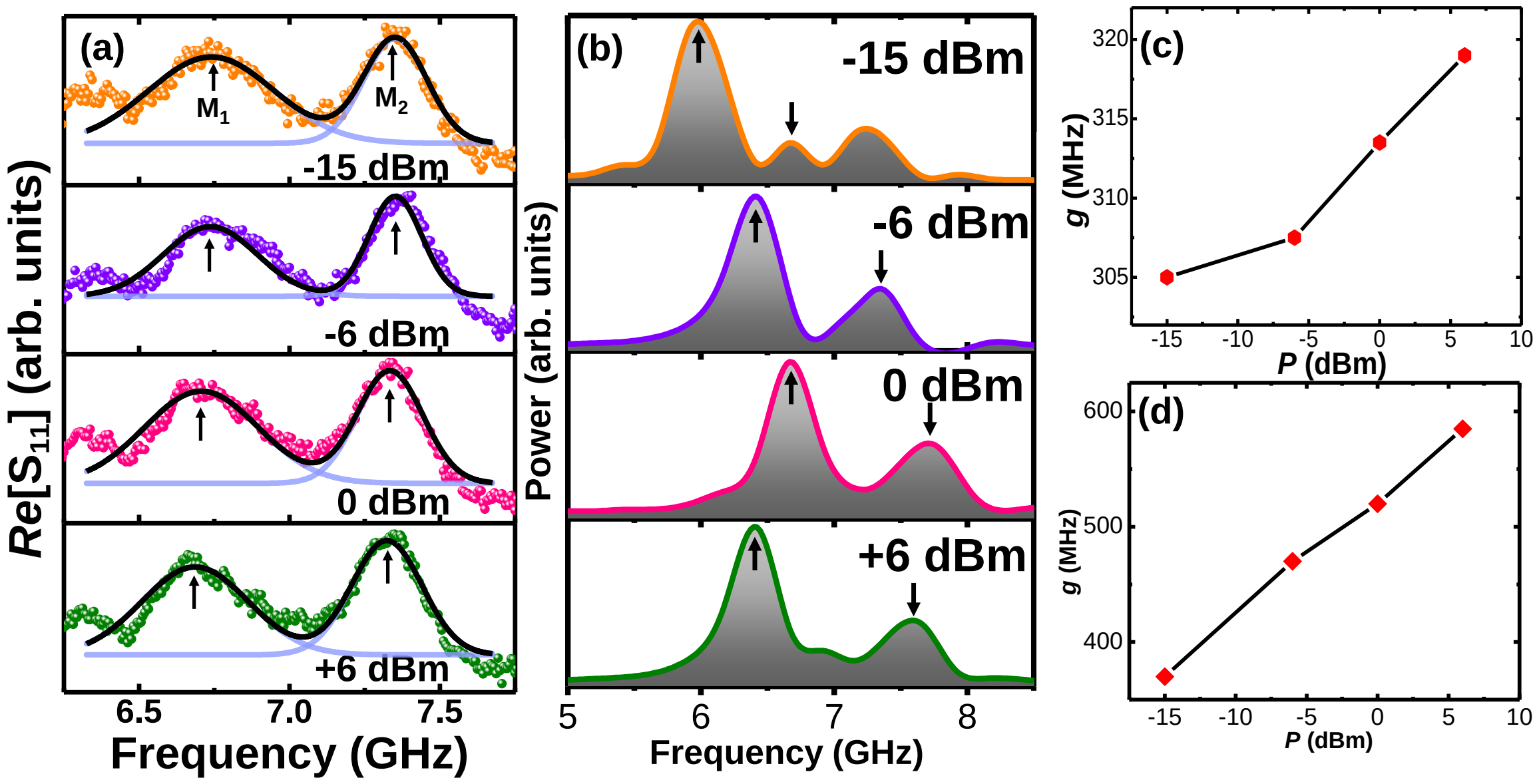

**Figure 4.** (a) The real component of the $S_{11}$ parameter is plotted against frequency for various power levels $P$ at $\varphi$ = 25°, highlighting the anticrossing behavior. The solid black lines correspond to theoretical fits, with the SW modes M1 and M2 indicated by upward arrows. (b) The corresponding simulated SW spectra are presented. (c) Experimental and (d) simulated values of the coupling strengths $g$'s are plotted as a function of $P$. In both cases, the solid lines are guide to eye.

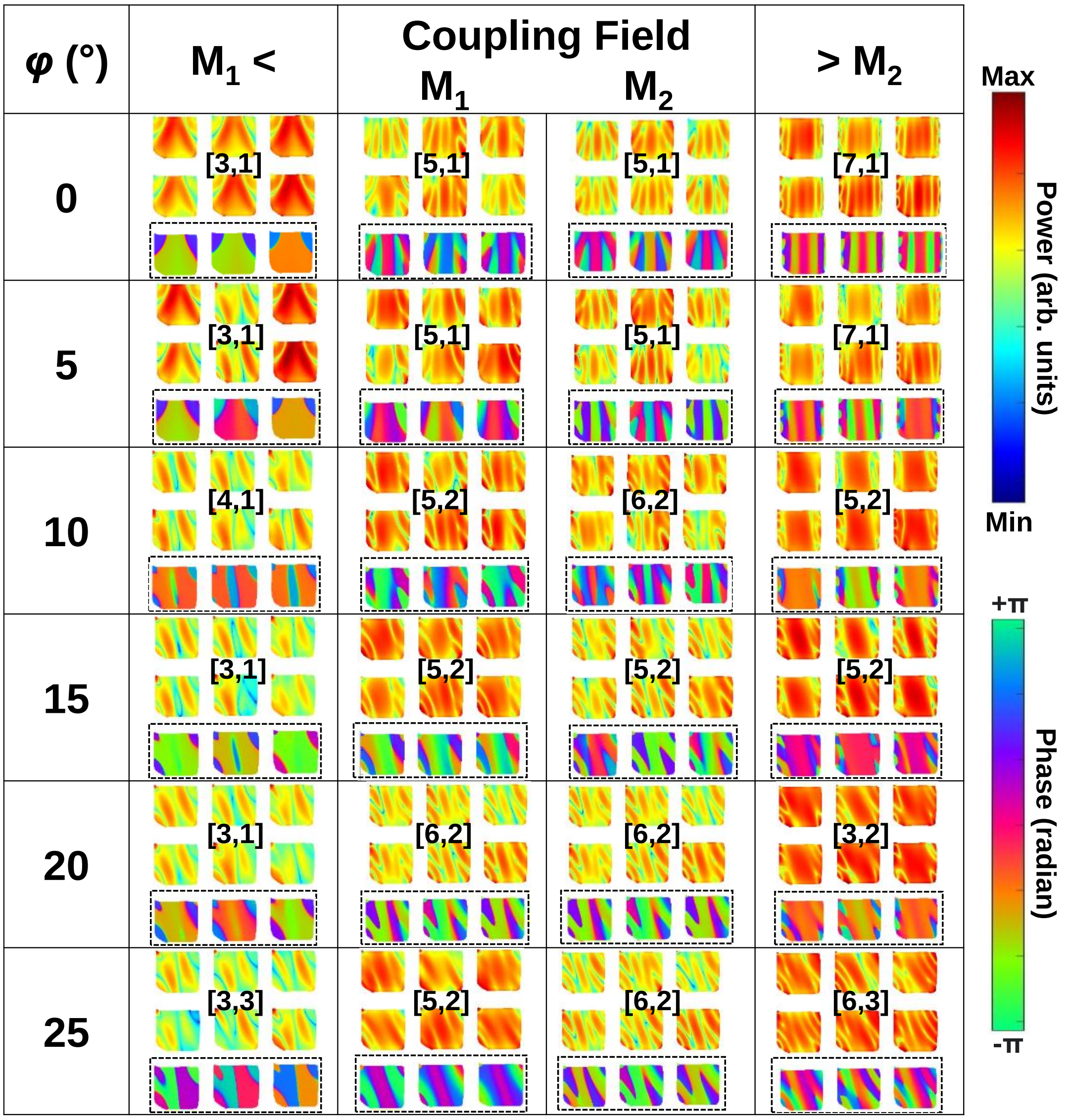

**Figure 5.** Simulated spatial distributions of power and phase of the coupled SW modes, both at and away from the anticrossing region, at various angular orientations of bias magnetic field. The quantization numbers for each SW mode are denoted as [m, n] in each scenario. The corresponding color maps for the power and phase profiles are provided on the right side of the figure.

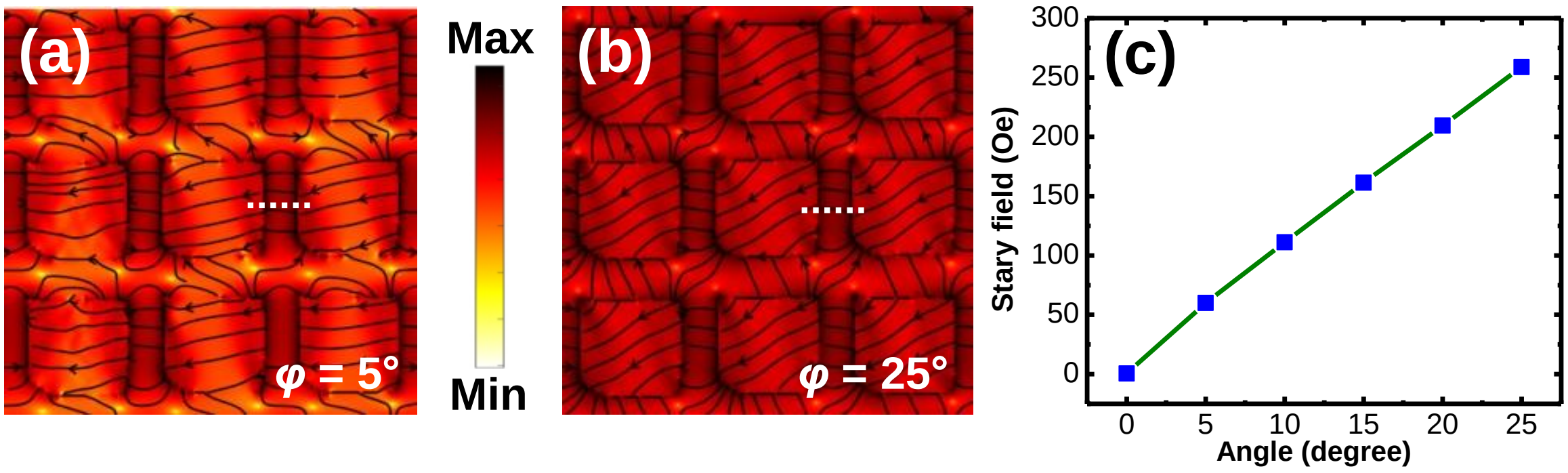

**Figure 6.** Simulated distributions of the magnetostatic field within the nanodot array at in-plane bias field angles of (a) $\varphi = 0°$ and (b) $\varphi = 25°$ at the anticrossing magnetic field. The corresponding color map is displayed between the two images. (c) The extracted magnetic stray field ($H_s$) at the midpoint between two nanodots in the central region of the array, derived from the line scans indicated by white dotted lines in (a) and (b), is plotted as a function of $\varphi$.

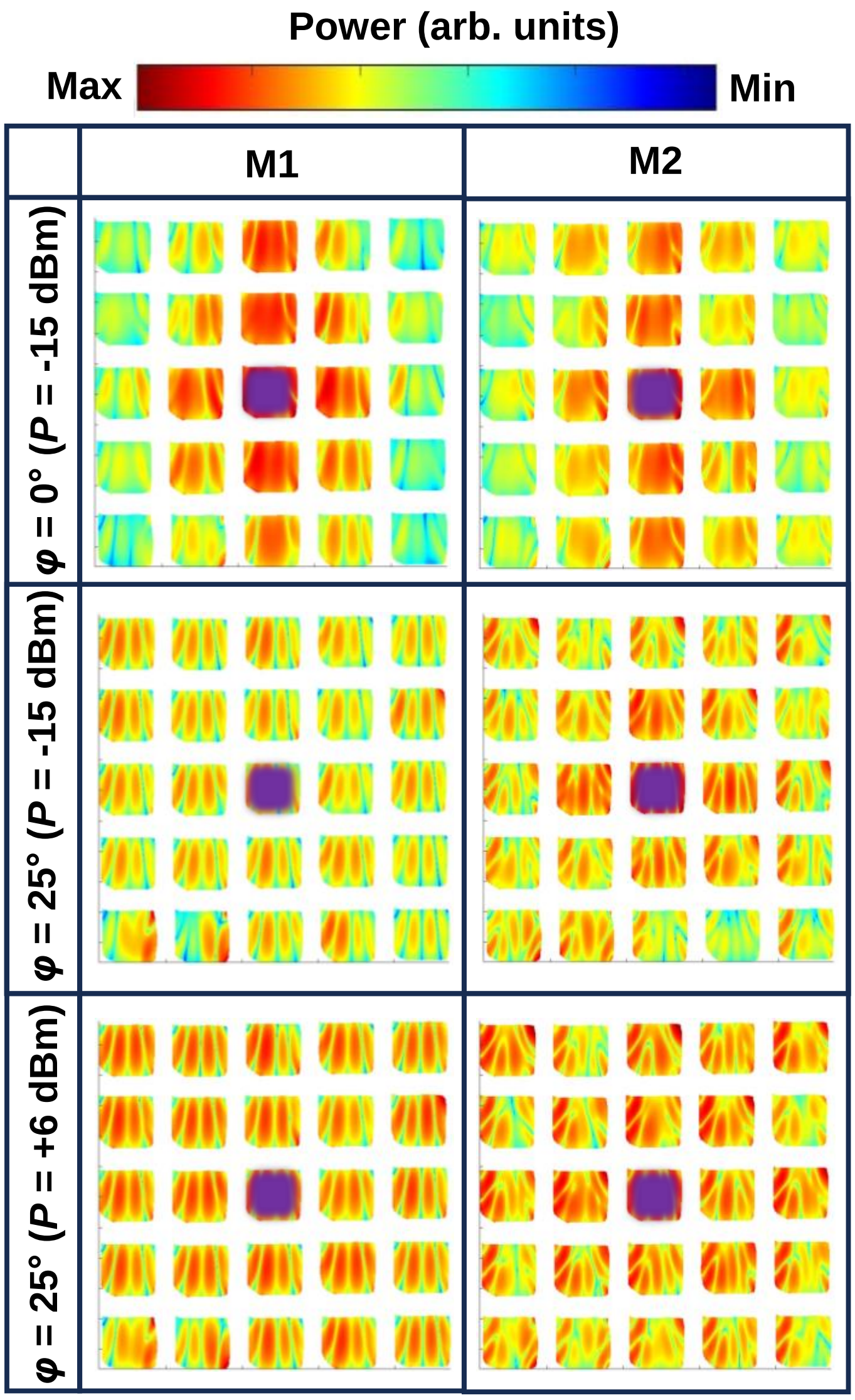

**Figure 7.** Simulated spatial distribution of power of the hybrid SW modes M1 and M2. The center of the array, highlighted in violet, was locally excited with a 'sinc' function having power levels of $P = -15$ dBm and $P = +6$ dBm at two specific angular orientations ($\varphi = 0°$ and $\varphi = 25°$) of bias magnetic field. The corresponding color map is displayed at the top of the figure.